\documentclass[traditabstract]{aa}
\usepackage{amsmath}
\usepackage{bm}
\usepackage{txfonts}
\usepackage{graphicx}
\usepackage{booktabs}
\usepackage{natbib}
\usepackage{astrojournals}
\usepackage[colorlinks=true,citecolor=blue]{hyperref}
\usepackage{epstopdf}

\newcommand{\RMS}[1]{\mathrm{RMS}\,[#1]}
\newcommand{\unit}[1]{\mathrm{#1}}
\newcommand{\unitp}[2]{\ensuremath{\mathrm{#1}^{#2}}}
\newcommand{\slfrac}[2]{\left.#1\middle/#2\right.}
\def\micron{\hbox{$\mu$m}}

\title{Phase Correction for ALMA with 183\,GHz Water Vapour Radiometers}

\author{B.~Nikolic\inst{1,2}  \and
        R.~C.~Bolton\inst{1,2} \and
        S.~F.~Graves\inst{1,2} \and
        R.~E.~Hills\inst{1,3} \and
        J.~S.~Richer\inst{1,2}   }

\institute{Astrophysics Group, Cavendish Laboratory, University of
  Cambridge, JJ Thomson Avenue, Cambridge CB3 0HE, UK \and
  Kavli Institute for Cosmology Cambridge, Madingley Road Cambridge,
  CB3 0HA, UK  \and
  Joint ALMA Observatory, Alonso de Cordova 3107, Vitacura, Santiago, Chile
  }

\abstract{Fluctuating properties of the atmosphere, and in particular
  its water vapour content, give rise to phase fluctuations of
  astronomical signals which, if uncorrected, lead to rapid
  deterioration of performance of (sub)-mm interferometers on long
  baselines. The Atacama Large Millimetre/submillimeter Array (ALMA)
  uses a 183\,GHz Water Vapour Radiometer (WVR) system to help correct
  these fluctuations and provide much improved performance on long
  baselines and at high frequencies. Here we describe the design of
  the overall ALMA WVR system, the choice of design parameters and the
  data processing strategy. We also present results of initial tests
  that demonstrate both the large improvement in phase stability that
  can be achieved and the very low contribution to phase noise from
  the WVRs. Finally, we describe briefly the main limiting factors to
  the accuracy of phase correction seen
  in these initial tests; namely, the degrading influence of 
  cloud and the residual phase fluctuations that are most likely to be
  due to variations in the density of the dry component of the air.}


\begin{document}

\maketitle

\section{Introduction}

The Atacama Large (sub-)Millimetre Array (ALMA) is an
aperture-synthesis telescope that will consist of 66 antennas
observing at millimetre and sub-millimetre wavelengths on
baselines ranging in length from 9\,m to 16\,km. The Earth's
troposphere has a large degrading effect on astronomical observations
at these wavelengths, and to minimise these effects ALMA is sited
at a high (5000\,m elevation) and very dry site in northern Chile.

Even at this site, the troposphere will have a significant effect on
most ALMA observations, consisting of two related phenomena \citep[see, for
example,][]{WilsonTools2010}:
\begin{enumerate}
\item Absorption of incoming astronomical radiation (and corresponding
  emission of incoherent thermal radiation).
\item Delay to the incoming astronomical radiation (i.e., non-unit
  refractive index) that is variable in time and in position.
\end{enumerate}
The two phenomena are physically related by the requirement of
causality \citep[see, e.g.,][]{1956PhRv..104.1760T} as mathematically
described by the Kramers-Kr\"onig relation. They are primarily caused
by the molecules of oxygen, nitrogen and water in the troposphere. The
degradation of the astronomical signals caused by the absorption is of
course irreversible, but this is not the case for the delay. If (as is
the case for ALMA) the individual elements of the interferometer are
smaller than the characteristic length-scale over which delay varies,
and the received signal is sampled faster than the characteristic
time-scale over which the delay changes, then the effects of variable
delay can \emph{in principle\/} be corrected in the data processing
step \citep[][]{1971MNRAS.154..229H}. For this to be possible, it is
however necessary to be able to estimate the delay variation due to
the atmosphere. In the case of ALMA a system of Water Vapour
Radiometers (WVR) operating at 183\,GHz is used for this purpose.

\subsection{Structure of the atmosphere}

Nitrogen, oxygen and other `dry' components of the troposphere are
well-mixed with each other and are in near pressure equilibrium. As a
result, the pressure and total column density of the troposphere
generally vary only slowly with time and position. In contrast, the
temperature of the dry air \emph{does\/} vary rapidly in time and
position. This is due to a combination of hydrostatic temperature
variation, localised heating and cooling at the Earth's surface and
the effect of wind-induced turbulence. These temperature fluctuations
have only a minor effect on the absorption by dry air but a
significant effect on the refractive index, leading to fluctuations in
apparent delay of astronomical signals. These delay fluctuations are,
in turn, the dominant cause of \emph{seeing} that impacts astronomical
observations at optical and infrared wavelengths from the Earth's
surface. These `dry' delay fluctuations due to temperature differences
of air also have a measurable effect at millimetre and sub-millimetre
wavelengths, but are generally much smaller than the effects of water
vapour.

Due to its large dipole moment, water vapour is a strong absorber
at millimetre and sub-millimetre wavelengths. It also significantly
increases the refractive index of air and therefore delays the
radiation, with one millimetre of precipitable water vapour
corresponding to an equivalent of approximately 6 millimetres of extra
`electrical path'. Unlike the `dry' air components, water vapour is
generally very poorly mixed with the other components of air, which
means that its concentration (or partial pressure) varies rapidly in
time and position. This in turn means that the integrated water vapour
along the line of sight of each element of an interferometer, and
consequently the apparent delay to each element, are fluctuating in
time in a way that is different for each element.

If we assume that the differences in concentration of water vapour are
driven by fully developed turbulence, then the likely difference in
concentration of water vapour between two points in the atmosphere is
an increasing function of the distance between these points up to some
`outer length scale'. \cite{1999RaSc...34..817C} show evidence that, at the VLA site in New Mexico, USA, this outer length-scale at which the
differences between two points no longer increase is at least as large as the maximum
baseline length of ALMA. This increase in the differences in water vapour
concentration with baseline length translates into an increase in the
magnitude of path fluctuations on longer baselines, making
measurements of astronomical signals less efficient and less accurate.

\subsection{Impact of phase errors on science data}
\label{sec:impact-phase-errors}

An aperture synthesis telescope is only sensitive to changes in the
difference of delays between pairs of elements. A fluctuating
difference in the delays leads to a fluctuation in the phase of the correlated signal (the `visibility') recorded for that pair.

Such phase fluctuations have a number of effects on the images of the
sky reconstructed from the visibilities \citep[see for
example][]{ALMANikolic582,1997A&AS..122..535L} two of which are most
important. Firstly, they cause decorrelation, i.e., a reduction of
the apparent amplitude of visibility recorded, because the signal is
not fully coherent for the duration of observation. The decorrelation
causes a loss of amplitude by a factor $\exp\left(-\frac{\phi_{\rm
      RMS}^2}{2}\right)$ where $\phi_{\rm RMS}$ is the
Root-Mean-Square fluctuation of phase. The magnitude of the noise in
the measurements is unaffected by these phase fluctuations and
therefore the decorrelation leads to a reduction in the sensitivity
of the telescope.

Secondly, the errors in the measured visibilities produce spurious
features on the maps that are produced.  These spurious features lead
to a further reduction in the dynamic range of the maps.  (The term
'dynamic range' is the ratio of the highest peak on the map to the
average noise in the parts of the map where there are no real
astronomical sources.)  Note that in addition to the errors in the
phases  caused directly by the atmospheric fluctuations, errors are
also introduced into the measured amplitudes because the amount of
decorrelation of each sample varies due to the random nature of the fluctuations.  Both sorts of error contribute to the spurious features in the map.

The magnitude of the atmospheric fluctuations increases on longer
baselines and, at millimetre and particularly sub-millimetre
wavelengths, a baseline length is reached where the path variations are a large fraction of a wavelength and no useful astronomical signal can be reconstructed. This means that, without correction of the phase fluctuations, the angular resolution of the telescope will be limited by the effects of the atmosphere.  The resolution achievable without correction is extremely variable, but a typical value is, coincidentally, similar to the optical seeing at good sites, i.e., around 0.5 arcseconds \citep[see for example the analysis of ALMA site-testing data by][]{ALMAEvans471}.

\subsection{Strategies for dealing with atmospheric phase
  fluctuations}

If no phase correction method is available, the only option is to limit the observing to
short baselines and/or to times when the atmosphere is sufficiently stable.
This is effective because on short enough baselines, the elements of
an interferometer look along similar lines of sight through the
atmosphere and therefore suffer from almost the same \emph{total\/}
path fluctuations; this means that the \emph{differential\/} path
fluctuation, which is what determines phase fluctuation of the
visibility, is small.  This strategy however directly limits the
maximum resolution attainable with the interferometer and reduces the amount of observing time available. 

Alternatively, if the surface brightness of the objects being observed is sufficiently high, it may be possible to use the
technique of self-calibration \cite[e.g.,
by][]{1984ARA&A..22...97P,1999ASPC..180..187C}. With this technique,
the observed interferometric visibilities are used to solve simultaneously
for the path errors to the antennas \emph{and\/} for the
reconstructed image of the sky. For this technique to be effective the signal-to-noise ratio in the time interval being solved for needs to be high enough to give phase errors  smaller than those caused by the atmosphere.  This technique works well at centimetre wavelengths, especially for compact non-thermal sources, but at millimetre and sub-mm wavelengths many objects will not have sufficiently high brightness temperatures to satisfy the
signal-to-noise requirement, especially on the longer baselines.  Several factors contribute to this increasing difficulty as one moves to shorter wavelengths.  These include the nature of the emission, which is generally thermal, the higher system noise temperatures and the fact that the timescales of the atmospheric fluctuations that cause significant decorrelation are shorter.  The large instantaneous bandwidth available provides some mitigation when the source has continuum emission, but that does not apply for objects where the emission is limited to a few spectral lines.

To alleviate the problems of low signal-to-noise ratio on the science
target, phase referencing (also known as fast-switching) can be implemented. In
this scheme a phase calibrator, nearby in the sky to the science
target, is observed frequently and the
phase measurements obtained for each antenna (from self-calibration)
are used to generate an interpolated phase correction table for the
source. This method is presented in \cite{1999RaSc...34..817C} and is
in routine use at the (J)VLA at 22\,GHz and higher frequencies.  For centimetre wavelengths, calibration on timescales of order one minute is effective, but at millimetre and especially sub-mm wavelengths much shorter calibration times are needed to prevent significant decorrelation.  Finding suitable phase calibrators that are sufficiently close to the target also becomes increasingly difficult at higher frequencies.

Another option is the paired-antennas method
\citep[see][]{1996RaSc...31.1615A} where sub-arraying (i.e.,
independent interferometric arrays operating side-by-side) is used so
that both the target source and a calibrator source are observed
simultaneously. Provided that the science target and calibrator are
close enough on the sky, phase measurements of the calibrator can be
used to determine the phase correction required above each of the
calibrator-array antennas and these can be transferred, directly or
with interpolation, to nearby antennas in the science-subarray.  This
technique has been used successfully at mm wavelengths at the CARMA
telescope \citep{2010ApJ...724..493P}.

Although they are effective at improving phase noise from the
atmosphere, both fast-switching and the paired-antennas methods come at a cost in terms of sensitivity, in the former case because time on source is reduced and in the latter because only a fraction of the array is available for science observing. In addition there are, in both cases, limitations on the accuracy that can be achieved.  In fast-switching, the non-continuous measurement of
the calibrator phase means that we are forced to interpolate in time
to estimate the phase corrections to apply to the observations of the science source, and the lines of sight through the atmosphere necessarily differ because the source and calibrator are not co-located on the sky. In the paired antenna scheme, whilst there is continuous measurement of calibrator phase, the path through the atmosphere is different for the science and calibrator sources, both because of the different sky positions and because of the different positions of the antennas on the ground.

\subsection{Radiometric phase correction}

It was realised early on in the development of interferometric telescopes operating at cm-wavelengths that water vapour fluctuations are an important limitation on achievable resolution
\citep{1138969,1971MNRAS.154..229H}. By this time the basic 
technique of using passive radiometers to do remote-sensing of the
water vapour content of the atmosphere was already developed
\citep{barrett_method_1962} and it was therefore soon proposed that
radiometers could be used to correct for the path delay due to water
vapour and therefore phase errors in an interferometer
\citep[e.g.,][]{Schaper1970ProcIEEE}.

The principle of radiometric phase correction is to make measurements of the sky brightness as a function of
time along the line of sight of the antennas and use these to infer the delays to the astronomical signal received at each antenna. Since water vapour is the largest contributor to the time-variable delays, radiometer
systems have most often been designed to observe at frequencies around strong water-vapour transitions. 

Most of the larger mm and sub-mm wave interferometric telescopes have either implemented or experimented with radiometric phase correction
techniques in order to improve the coherence of the astronomical
signal on longer baselines. Systems based on the 22\,GHz water vapour
line have been installed or experimented with at the IRAM Plateau de Bure
Interferometer (PdBI) in the French Alps \citep{2002ASPC..266..238B},
at the NRAO Very Large Array in New Mexico and at the Owens Valley Radio
Observatory in California \citep{2000ASPC..217..317W}. Only the system
at PdBI is in current operation.  22\,GHz radiometers are also used on
some VLBI telescopes, for example at Effelsberg
\citep{2007astro.ph..3066R}.

A broadband continuum radiometric phase correction technique where the
atmospheric emission away from specific transitions, but which is still
mostly due to water vapour, has been tried at
Berkley-Illinois-Maryland Array \citep{zivanovic_new_1995}, IRAM PdBI \citep{Bremer1995} and at the Sub-Millimeter Array \citep{2004ApJ...616L..71B}.

Finally, phase correction using the 183\,GHz water vapour line has
been demonstrated at the James Clerk Maxwell Telescope -- Caltech Submillimeter
Observatory interferometer \citep{2001ApJ...553.1036W}. The ALMA
radiometer system is to a large extent based on that concept.

\section{System overview}

The ALMA specification, at system level, for correction of phase
fluctuations due to water vapour is given by:
\begin{equation}
  \delta L_{\rm corrected} \leq
  \left(1+\frac{c}{1\,\unit{mm}}\right)\,10\,\unit{\micron} +
  0.02 \times \delta L_{\rm raw}
\label{eq:almaspec}
\end{equation}
where the symbols have following meanings:
\begin{description}
\item[$\delta L_{\rm raw}$] is the uncorrected path fluctuations on a
  baseline
\item[$c$] is the Precipitable Water Vapour (PWV) column along the
  line of sight of an antenna
\item[$\delta L_{\rm corrected}$] is the specification for the
  per-antenna path fluctuations after the correction.
\end{description}
The quantity normally measured is the phase fluctuation on a baseline,
which is allowed to be up to $\sqrt{2}$ larger than the per-antenna
specification.  The specification in Eq~(\ref{eq:almaspec}) applies on
timescales shorter than 180\,seconds; the specification assumes that
any residual errors on timescales longer than 180\,seconds will be
adequately corrected by observations of point source phase calibrators
near to the science target.

If these specifications are just met, they imply a 7\% amplitude loss
due to decorrelation (see discussion in Section~\ref{sec:impact-phase-errors}) from residual phase
errors if observing at $\lambda=350\,\micron$ in 0.5\,mm PWV; 2.5\%
loss if observing at $\lambda=850\,\micron$ at 1.1\,mm PWV; and less
than 1\% loss when observing at $\lambda=3\,\unit{mm}$ when the line
of sight PWV is 3\,mm. By comparison, the loss due to atmospheric
absorption when observing at $\lambda=350\,\micron$ through 0.5\,mm
PWV is about 50\%, i.e., much larger than the loss due to
decorrelation.  The losses due to decorrelation are however highly variable and hard to estimate.  As a result they affect both the image quality and the flux calibration accuracy.  By contrast, the losses due to absorption need not have a significant effect on imaging or accuracy, apart from the inevitable loss of sensitivity, so long as appropriate calibration techniques are applied. 

The ALMA phase correction system consists of 183\,GHz WVRs installed on all
of the 12\,m diameter antennas. The WVRs measure the
observed sky signal, integrate the signal for an operator-selected time interval and apply an internal calibration, so that the output values are in units of antenna temperature. 
The integration times used are typically around one second, as this
corresponds to characteristic minimum timescale ($\sim D/u$ where
$D=12\,\unit{m}$ is the antenna diameter and
$u\sim10\,\unit{m}\,\unitp{s}{-1}$ is the wind speed) over which the
observed signal varies.  The WVR measurements are collected by the
ALMA correlator subsystem, recorded to the permanent data store and
optionally used for on-line phase-correction.

\section{WVR Instrument Design}

\begin{figure}
  \includegraphics[clip,width=0.99\columnwidth,trim=80 370 30 120]{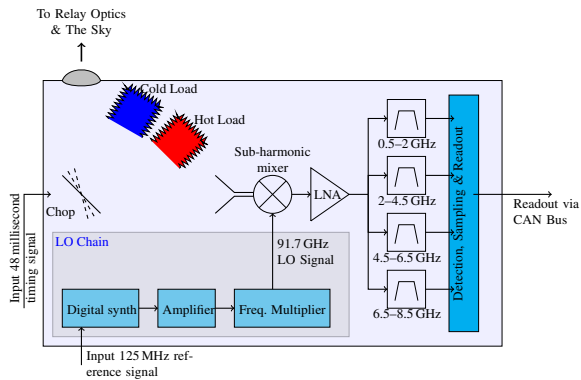}
  \caption{Conceptual design of the water vapour radiometers}
  \label{fig:wvrconcept}
\end{figure}

A block diagram of the conceptual design of the radiometers deployed
by ALMA is shown in Figure~\ref{fig:wvrconcept}. This is an evolution
of the design originally described by \cite{2001ApJ...553.1036W} which
was used as a starting point for two ALMA prototype WVRs designed and
built by a collaboration between the University of Cambridge and
Onsala Space Observatory. These prototypes were in turn used as input
for the detailed design and production of the final production WVRs by
Omnisys Instruments AB, Sweden. The detailed design of production
units is described by \cite{Tera20Emrich}.  In this section we
describe the design parameters of the WVRs and their justification.

\subsection{Frequency response}

The objectives of the frequency response and filter design of the WVRs
were:
\begin{itemize}
\item Maximum sensitivity to water vapour fluctuations over a range of
  total water vapour column conditions.
\item Sufficient frequency resolution to avoid using the saturated
  part of the line when conditions are wet and to avoid the wings when
  conditions are dry.
\item Measurement of both the inner part of the water vapour line and
  the wings to allow inference of atmospheric properties such as total
  water vapour column and also constraints on its vertical distribution.
\item Keep the design as simple as possible and limit the costs.
\end{itemize}

To maximise the sensitivity, the entire region of the spectrum around
the 183\,GHz water vapour line is sampled simultaneously using a
Double-SideBand (DSB) mixing system with four fixed Intermediate
Frequency (IF) filters. The alternative arrangement of using a single
filter, sideband separation and tunable Local Oscillators (LOs) to
scan a single detection band across the line is used in some
commercial atmospheric sounders but would lead to extra complexity and
lower sensitivity (as only a part of the spectrum is sampled at any
one time) while the additional frequency resolution is not thought to
be an advantage in our application.

During the design and prototyping stages the effects of clouds were
recognised as potential sources of errors in the WVR phase correction
technique. In order to try to reduce this problem, a sideband-separation design leading to a
dual-single-sideband (2SB) system was considered and implemented in one
of the prototypes.  Such a 2SB system can distinguish a sloping
continuum spectrum from one which is symmetric around the LO frequency and can 
therefore be used to separate the contribution to the sky
brightness from water vapour and from cloud droplets (which has
opacity that approximately varies as $\nu^2$).  The extra complexity
and cost of sideband separation was however judged to outweigh the
benefits and this option was therefore descoped for the ALMA production radiometers.

\begin{figure}
  \includegraphics[clip,width=0.99\columnwidth]{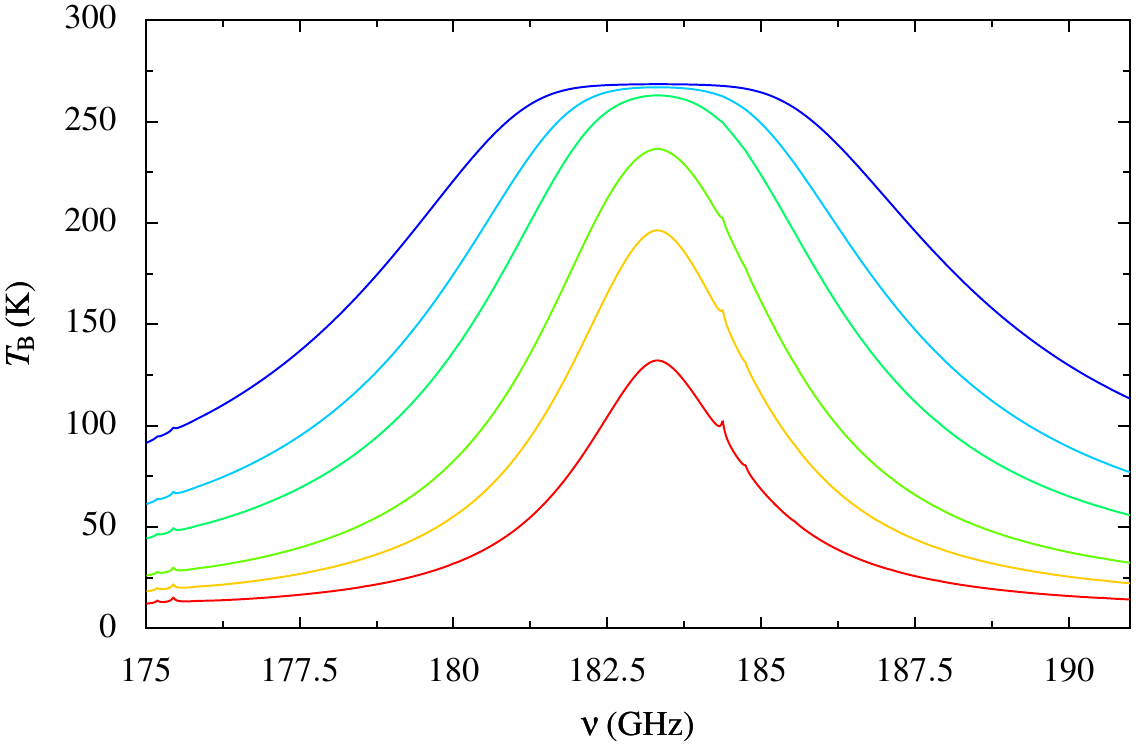}
  \caption{Model brightness of the atmosphere at frequencies around
    the 183\,GHz water vapour line for six values of PWV along the
    line of sight: 0.3, 0.6, 1, 2, 3 and 5\,mm (from lowest red line
    to highest blue line). The model was computed using the ATM
    program by \cite{Pardo2001}, using the source code the model as
    used by ALMA (this is available for public download at
    \url{http://www.mrao.cam.ac.uk/~bn204/alma/atmomodel.html}). }
  \label{fig:wvline}
\end{figure}

\begin{figure}
  \includegraphics[clip,width=0.99\columnwidth]{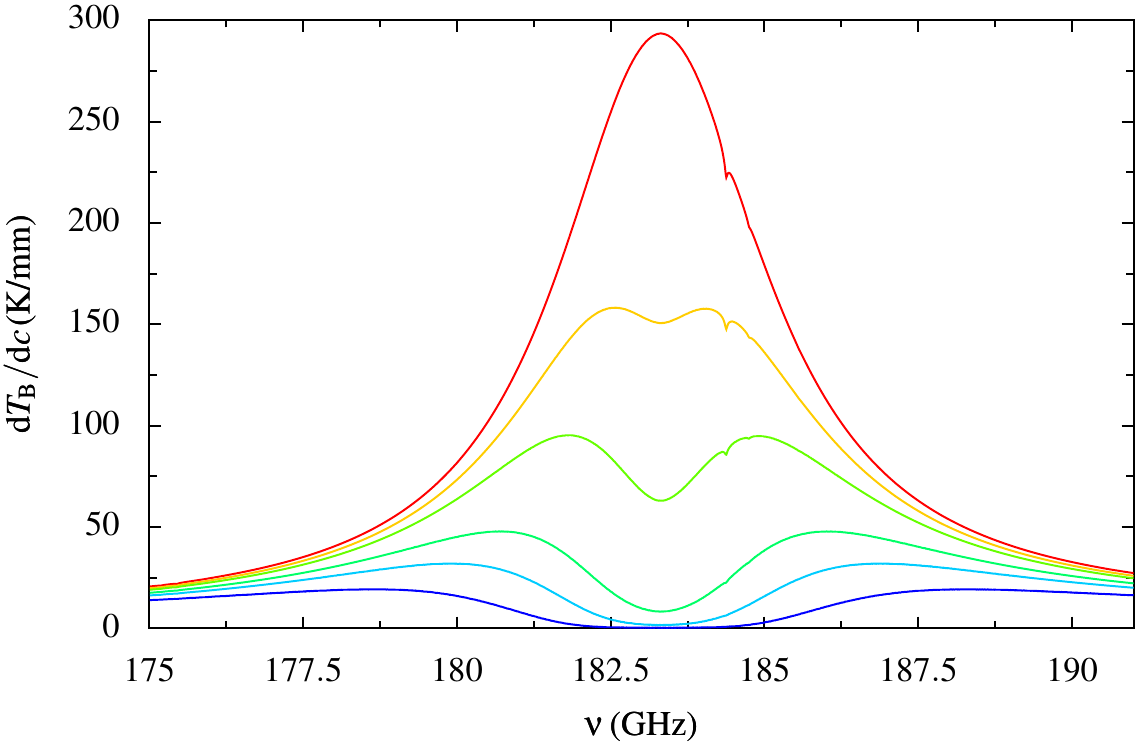}
  \caption{Differential of brightness of the atmosphere with respect
    to the total line of sight precipitable water vapour ($c$),
    computed for the same six values of total PWV as in
    Figure~\ref{fig:wvline}.}
  \label{fig:wvdtdc}
\end{figure}

\begin{figure}
    \includegraphics[clip,width=0.99\columnwidth]{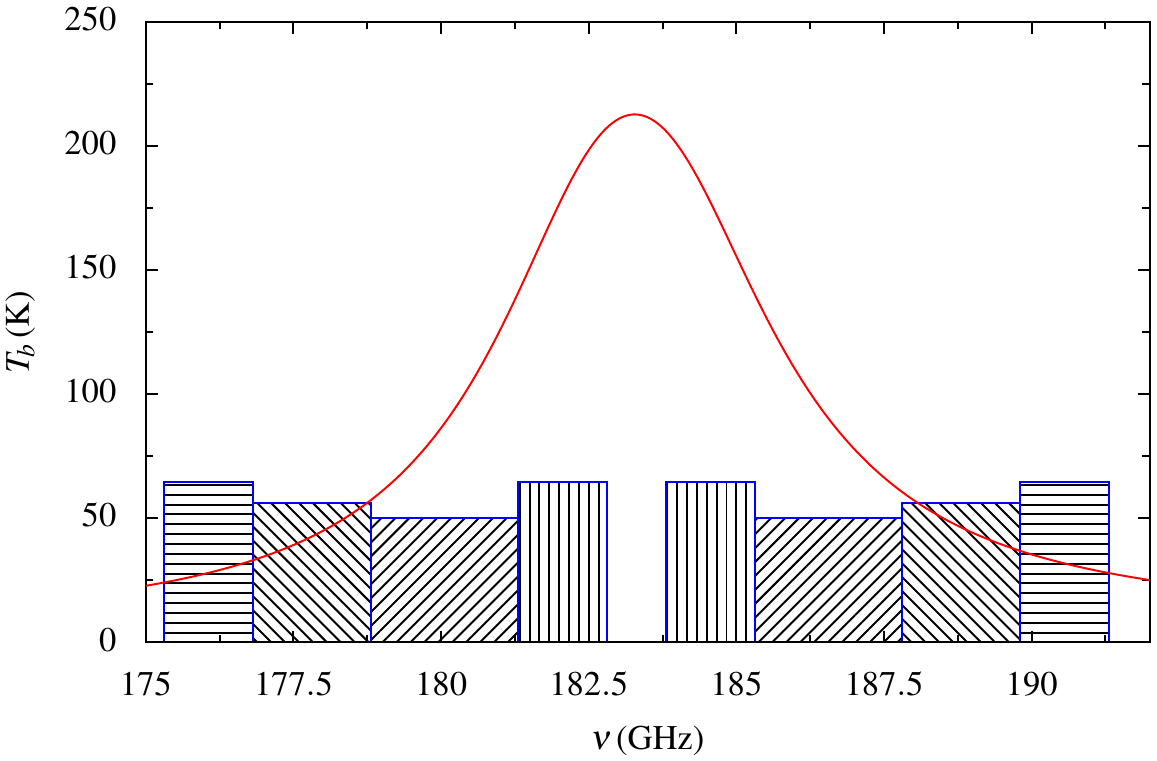}
    \caption{The design band-passes of the four channels of ALMA WVRs,
      together with a plot model brightness temperature for 1\,mm of
      PWV.}
    \label{fig:filters}
\end{figure}

Selection and optimisation of the filters frequencies and bandwidths
is discussed in detail by \cite{ALMAHills568}. The main conclusions of this optimisation study are that:
\begin{itemize}
  \item For optimum sensitivity it is best to divide the entire
    available IF bandwidth between the filters.
  \item Estimates of the continuum (which is dominated by clouds when they are 
    present) are most constrained by the outermost channel and
    therefore this channel should be kept relatively narrow so that is contains only a small contribution from the wing of the line.
\end{itemize}

A model computation of sky brightness around the 183\,GHz water vapour
line is shown in Figure~\ref{fig:wvline} for a range of total water
vapour columns that is representative of typical conditions at the
ALMA site. It can be seen that the conditions range from optically
thin at the centre of the line to completely optically thick in a
range of around 2\,GHz around the centre of the line. This can also be
seen Figure~\ref{fig:wvdtdc}, which shows the rate of change of sky
brightness with respect to changes of the total water vapour column
over the same frequency range as Figure~\ref{fig:wvline}. Higher
values of this rate of change mean that for the same magnitude of
noise in the WVR receiver system, a smaller phase noise is achievable
in the corrected data. It can therefore be seen from this plot that
the optimum frequencies for measurement of water vapour fluctuations
range from the centre of the line (under very dry conditions) to about
to about 5\,GHz away from the centre (for about 5mm of water). In
order to allow for a channel which measures primarily the continuum
contribution, the IF bandwidth selected was 0.5\,GHz to 8.0\,GHz
divided into four channels. The final design of the filter centres and
bandwidths is illustrated in Figure~\ref{fig:filters}, together with a
model computation of sky brightness of the 183\,GHz water vapour line.

\begin{figure}
  \includegraphics[width=0.99\columnwidth]{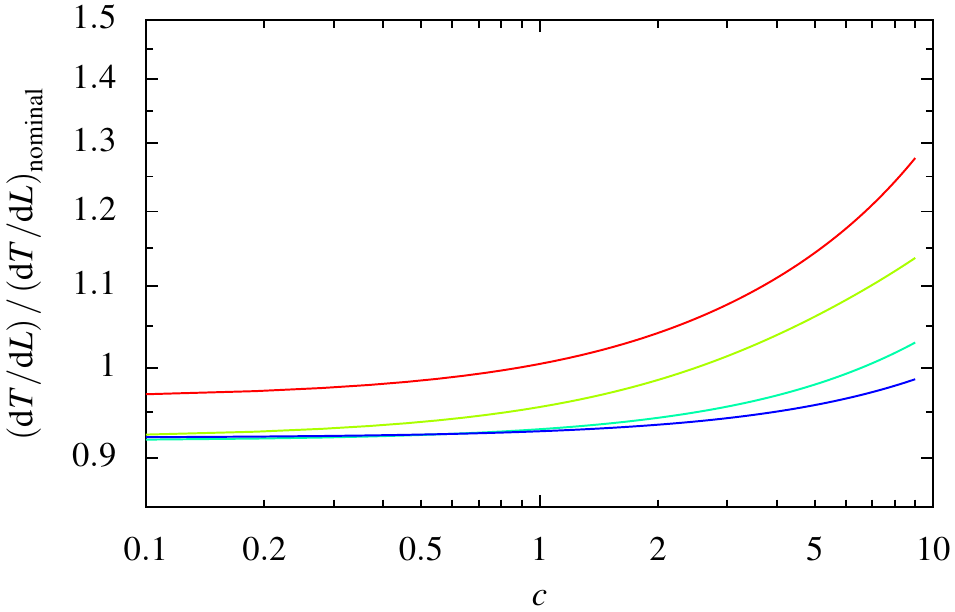}
  \caption{Relative change in the model phase correction
    coefficients for a 5\,\% shift in the centre frequency of each
    filter versus the total column
    of water vapour in mm. Red to blue lines (following the spectrum) are channels 1
    to 4.}
  \label{fig:filtersens}
\end{figure}

Since mass-production of WVRs with frequency responses that are very
uniform is difficult, ALMA adopted a manufacturing specification with
a 5\% tolerance on the centre frequency and bandwidth of the response
of the entire system.  To obtain the required phase-correction
performance it is however necessary to have better knowledge than this
of the actual response.  This is illustrated by
Figure~\ref{fig:filtersens} which shows a model computation of the
change of the constant of proportionality between sky brightness
fluctuations and path fluctuations for a 5\,\% change in the channel
centre frequency. It can be seen in this figure that a
mis-characterisation of the centre frequency of that magnitude can
easily lead to errors of a few to 10 per cent in the phase correction
coefficients, which will typically lead to similar errors in the
derived path correction. Such errors would be significant when
compared to the specification shown in
Equation~(\ref{eq:almaspec}). For this reason the frequency response
of each ALMA WVRs was characterised by the manufacturer after final
assembly to better than 1 percent accuracy in the centre frequency and
bandwidth of each channel.

\subsection{Sensitivity requirements}
\label{sec:sens-requ}
\begin{figure}
  \includegraphics[width=0.99\columnwidth]{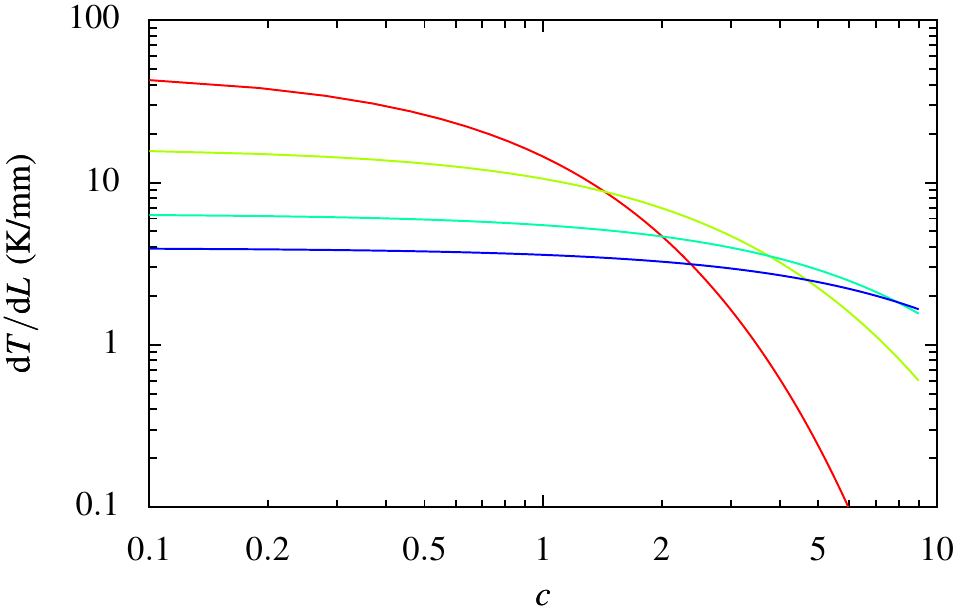}
  \caption{Model phase correction coefficients versus the total column
    of water vapour. Red to blue lines (following the spectrum) are
    channels 1 to 4, i.e., innermost to outermost channel. Note that
    the units of the ordinate are Kelvin per millimetre of extra path
    in contrast to Figure~\ref{fig:wvdtdc} which has units of Kelvin
    per mm of precipitable water vapour. }
  \label{fig:dtdlssimple}
\end{figure}

A basic limit on the accuracy of phase correction arises from the
white-spectrum thermal-like noise arising in the mixer and amplifier
of the WVR receivers. The magnitude of the errors introduced by this thermal-like noise depends on the atmospheric
conditions (primarily the total water vapour column), the integration time
used and the way in which the signals for the four channels of the
WVRs are combined. The thermal-like noise fluctuation in the measured
sky brightness is given by:
\begin{equation}
\delta T_{\rm ideal}=2 T_{\rm sys}/\sqrt{Bt}
\label{eq:radiometereq}
\end{equation}
where: 
\begin{description}
  \item[$T_{\rm ideal}$] is the fluctuation, labelled `ideal' because it
    does not take into account effect the gain fluctuations discussed
    in Section~\ref{sec:gain-absol-calibr},
  \item[$B$] is the bandwidth,
  \item[$t$] is the total integration time,
  \item[$T_{\rm sys}$] is the system noise temperature (and is of
    the order of 1000\,K for ALMA production systems).
\end{description}
The factor of two in Equation~(\ref{eq:radiometereq}) arises because
the radiometers measure the sky for only about half of the time while
the calibration loads are measured in the other half (this gives a
factor of $\sqrt{2}$); and, because the final measurement is derived
approximately from the \emph{difference} of measurements on the sky
and the calibration loads (giving another factor of $\sqrt{2}$). A
lower-noise design is possible by having two separate receivers
arranged so that one is observing the calibration loads while the
other is observing the sky and vice versa. This was implemented on the
prototype WVRs but was not adopted for the production requirements for
reasons of cost and complexity.

The relationship between a small change in sky brightness and the
consequent change in the electrical path to the antenna is given by
`phase-correction coefficients' which we denote by $\slfrac{{\rm d}  L}{{\rm d} T_{{\rm B}}}$.
Plots of phase-correction coefficients computed from models for the
four channels of the production WVRs and for a wide range water vapour
columns are shown in Figure~\ref{fig:dtdlssimple}. The critical region
of PWV for sensitivity is around 2\,mm of PWV where the line begins to
saturate and the effective sensitivity of the central channel begins
to decrease rapidly. The phase correction coefficients at around 2\,mm PWV are around 4--6$\,\unit{K}\,\unitp{mm}{-1}$ while the
specification in Equation (\ref{eq:almaspec}) calls for a path error
of 0.03\,mm.  This calculation indicates that if the noise in
measuring the sky brightness can be held to of order 0.1\,K, the
specification can be met in this critical region while still using
data from only one channel. A more detailed analysis of this topic is given by \cite{ALMAHoldaway515}.

\subsection{Gain and absolute calibration}
\label{sec:gain-absol-calibr}

In order to achieve a high effective sensitivity, the WVRs must be
calibrated frequently enough to remove the effect of gain
fluctuations. These fluctuations in gain usually have a $1/f$
spectrum. Additionally, WVRs for ALMA need to have a good absolute
calibration, since the total power of sky emission is used infer the
water vapour column and other atmospheric parameters.  They also need
good stability in time and with respect to small motions in the
antenna.

For a standard Dicke radiometer, an estimate of the effective noise including the effect of gain fluctuations is:
\begin{equation}
  \delta T_{\rm src}= \sqrt{\delta T_{\rm ideal}^2 + 
                           \left[(T_{\rm src}-T_{\rm ref})\frac{\delta
                               G}{G}
                           \right]^2}
\end{equation}
where:
\begin{description}
   \item[$\delta T_{\rm ideal}$] is the fluctuation
  of an `ideal' radiometer as discussed in Section~\ref{sec:sens-requ}
  \item[$T_{\rm src}$] is the effective temperature of the source being measured, i.e. the sky brightness 
  \item[$T_{\rm ref}$] is the effective temperature of reference, i.e. the calibration load or loads
  \item[$G$] is the gain of the radiometer
  \item[$\delta G$] is the fluctuation in the gain 
\end{description}
Normally a Dicke-switched radiometer is designed so that $T_{\rm ref}$
is close to $T_{\rm src}$ and the therefore the contribution of gain
fluctuations to the uncertainty of measurement is minimised. This is
however hard to achieve in the case of ALMA's 183\,GHz WVRs because:
\begin{enumerate}
\item The sky brightness varies strongly as a function atmospheric
  conditions, frequency and airmass, with the range corresponding to
  brightness temperatures from 25\,K to about 260\,K. Therefore no
  single reference load temperature can be suitable for all
  observations.
\item It is considerably cheaper and more reliable to have reference
  loads near the ambient temperature, whereas the sky brightness is always
  smaller than these ambient temperatures.
\end{enumerate}
For these reasons the ALMA WVRs operate as \emph{unbalanced\/} Dicke
radiometers where the differencing scheme removes some but not all of the effects of gain fluctuations, and some form of
frequent calibration for gain fluctuation must be used.  This is
implemented by observing \emph{two\/} calibration loads, at different
temperatures, $T_{\rm cold}$ (about 280\,K for the ALMA production
WVRs) and $T_{\rm hot}$ (about 360\,K).

The gain is estimated as:
\begin{equation}
  \hat{G}=\frac{V_{\rm hot}-V_{\rm cold}}{T_{\rm hot}-T_{\rm cold}}
\end{equation}
where $V_{\rm cold}$ is the signal when observing the cold load and
$V_{\rm hot}$ is the signal when observing the hot load. Therefore:
\begin{align}
  T_{\rm src}&= \frac{V_{\rm src}-V_{\rm ref}}{\hat{G}} + T_{\rm ref}\\
  V_{\rm ref}&=\frac{V_{\rm hot}+V_{\rm cold}}{2}\\
  T_{\rm ref}&=\frac{T_{\rm hot}+T_{\rm cold}}{2}
\end{align}
but $\hat{G}$ now includes fluctuations due to measurement error of
$V_{\rm hot}$ and $V_{\rm cold}$. In fact unless $({T_{\rm hot}-T_{\rm
    cold}})$ is much greater than $(T_{\rm src}-T_{\rm ref})$ then $\hat{G}$
must be smoothed. If the gain is not smoothed, then even without any gain
fluctuations the effective noise will be dominated by the errors in
the gain estimate rather than the gain fluctuations themself. An
approximate relationship is:
\begin{equation}
  \delta T_{\rm src} \approx  \delta  T_{\rm ideal}
  \sqrt{1+\frac{2}{t_{\rm S}}\left(\frac{T_{\rm src}-T_{\rm ref}}{T_{\rm hot}-T_{\rm
    cold}}\right)^2}
\end{equation}
where the factor of two arises because the calibration loads are each
observed for only half of the time that the sky is observed and
$t_{\rm S}$ is the smoothing factor, which is the ratio of the time over which the gain is smoothed to the integration time used for the measurement of the sky brightness temperature.  For the case of ALMA WVRs, a smoothing factor of at least 60 is required to ensure that the errors in the gain estimate are not dominant.

In fact, for unbalanced Dicke-switched radiometers like this, an
alternative gain estimation scheme is possible which makes use of the facts that:
\begin{enumerate}
  \item The receiver noise temperature and any additive post-detection offsets are relatively stable in time.
  \item The \emph{total\/} signal when observing a calibration load can then be used
    to calibrate the gain, i.e.:
    \begin{equation}
      \hat{G'}\sim \frac{V_{\rm ref}}{T_{\rm ref}+T_{\rm rec}}.
    \end{equation}
  \item Longer time scale changes in the gain and receiver
    temperature/offset can be estimated from two-load observations
    using ample smoothing.
\end{enumerate}
This has the advantage that, as denominator ${T_{\rm ref}+T_{\rm
    rec}}$ is large, the significance of measurement fluctuations of
$V_{\rm ref}$ is small and no smoothing on gain needs to be
employed, enabling correction of shorter timescale gain fluctuations.
This scheme was explored for the use on the ALMA WVRs but was found not to be necessary.

The overall absolute calibration of the WVRs is fixed by observing
external ambient and liquid nitrogen calibration loads during the
in-factory characterisation of the units.  The absolute calibration is
important as the absolute sky brightness is used to fit an atmospheric
model and to estimate the quantities needed for translation of WVR fluctuations into path changes. 
 
\subsection{Optical design}

To ensure accurate phase correction it is important that the
radiometer and astronomical beams overlap closely so that the
radiometer is measuring the emission from the same water vapour which
is delaying the astronomical radiation.  This was achieved at ALMA by
mounting the WVRs in the focal plane of the antennas so that they use
the same primary and secondary mirror optics as the astronomical
receiver. In fact the WVR beam corresponds to the bore-sight of the
telescope, with the four astronomical beams closest to it being the
highest frequency astronomical receivers (Bands 7, 8, 9 and 10) since
these are the ones for which the phase correction is most
important. The angular separation between the WVR beam and
astronomical beams is between 3.5 and 9 arc-minutes, corresponding to
a divergence of between 1 and 3 metres at 1000\,m above ground
level. At this distance from the antennas the beams are still of order
12m across, so this separation corresponds to about 10\%--30\% of the
beam width. The quantitative impact of this beam divergence is
investigated by \cite{ALMANikolic573}.  Fixed relative pointing
between astronomical and WVR beams is guaranteed by rigid mounting of
the radiometer on the Front-End Support Structure that also supports
the astronomical receivers.

The sensitivity of the WVRs is high enough that they can be quite
sensitive to spill-over radiation, especially if it depends on time or the pointing of the antenna. For example, a 0.1 percent change in the spill-over
onto the ground will lead to about a 0.3\,K change in the observed signal,
which is more than 3 times higher than the thermal noise in a one-second
integration time.  Such changes in spill-over would be confused with real changes of sky brightness and therefore be translated into inaccurate corrections of the phase of the astronomical signal. 

The magnitude of the spill-over for ALMA WVRs is minimised by
under-illuminating the optics.  The overall antenna gain of
the WVR system is not important for WVR operation since, to first order, the system needs only to measure the \emph{surface brightness\/} of sky emission along the boresight of the antenna.  Therefore the WVR system was designed to illuminate the primary reflector
with an edge taper of -16dB and a maximum of 2\% spill-over past the
secondary. The under-illumination, together with adequate sizes for 
the relay optics between the WVR and the secondary mirror, means that the
overall forward efficiency of the system is over 95\%. The
under-illumination does mean that the volume of the atmosphere that
affects the astronomical signal is a little larger than that measured by the radiometer but this is expected to have only a small effect on the final results \citep[see][]{ALMANikolic573}.

\section{Data processing}

As described above, the WVR measurements are calibrated internally in the radiometer and are
output as sky brightness in units of Kelvin. The subsequent data
processing uses these measurements to correct the astronomical data for the effects of the atmosphere. ALMA has two software implementations of this processing:
an on-line system which applies the phase correction, in the software associated with the correlator, before the visibilities are stored; and an off-line system which phase-rotates the visibilities during the standard interferometric data reduction.  The on-line system is part of the overall on-line telescope calibration system `TelCal' described
by \cite{2011ASPC..442..277B}, while the off-line system is the
separate program `{\tt wvrgcal}' described by \cite{ALMANikolic593}
and which is now incorporated into the
CASA\footnote{\url{http://casa.nrao.edu/}} distribution. The
principles of the two data processing implementations are broadly
similar but where they are different the description below is of the
off-line system.

At frequencies around
183\,GHz the relationship between sky brightness and water vapour column (and
therefore excess path) is highly non-linear because the water vapour line is
close to saturation even in very dry conditions. This is in contrast to
the situation at 22\,GHz, where the line is relatively weak and
therefore not saturated, but where other effects such as variable
spill-over and cloud can become very important.  This non-linearity,
together with the dependence of the water vapour line on temperature and
pressure of the atmosphere, means that a multi-dimensional non-linear
model must be fitted to the observed sky brightness to determine the
\emph{total\/} excess path.

For ALMA phase-correction we are however primarily interested in the
change of the electrical path to the antenna over small changes of
direction on sky and for small periods of time. For small enough
changes, the path can be linearised so that $\delta L \approx
\left(\slfrac{{\rm d} L}{{\rm d} T_{{\rm B},k}}\right) \delta T_{{\rm B},k}$ where
$\delta T_{{\rm B},k}$ is the fluctuation in the $k$-th radiometer
channel and each channel produces an independent estimate of the path
fluctuation. These independent path estimates can then be combined to obtain a single best estimate:
\begin{align}
  \delta L \approx \sum_k w_k \frac{{\rm d} L}{{\rm d} T_{{\rm B},k}} \delta
      T_{{\rm B},k} 
\label{eq:phase-corr-rel}
\end{align}
where $w_k$ is the weight assigned to each channel.  The sum of the weights is unity.

The data processing steps consist of:
\begin{enumerate}
 \item Estimating the phase-correction coefficients $\slfrac{{\rm d}
     L}{{\rm d} T_{{\rm B},k}}$.
 \item Choosing how to combine the four channels, i.e., choosing a set
   of weights $w_k$.
 \item Optionally filtering the observed fluctuations $\delta
   T_{{\rm B},k}$ to reduce the effect of noise in radiometers.
 \item Applying the phase correction to the observed visibilities.
 \end{enumerate}

 In the present implementation of phase correction software, the
 coefficients $\slfrac{{\rm d} L}{{\rm d} T_{{\rm B},k}}$ are
 estimated by taking the values of the observed sky brightness in the
 four WVR channels for one integration time in the middle of the
 astronomical observation, and fitting a model for the atmospheric
 emission. The model used is a relatively simple, thin
   single-layer model with pressure, temperature and column density of
   the water vapour all treated as free parameters constrained to a
   range of feasible values. The model and the procedure used to fit
   it to the observations are described in detail by
   \cite{ALMANikolic587} and \cite{ALMANikolic593}. The same model,
   together with the inferred atmospheric parameters, is then used to
   estimate the phase correction coefficients. We show in
 Section~\ref{sec:phase-corr-results} that this estimate of the phase
 correction coefficients is sufficiently good for the longest baseline
 lengths that we have been able to test so far ($B_{\rm max}\sim
 600\,\unit{m}$). A fit to observations from just one antenna is used to 
calculate the phase correction coefficients for the entire 
array. We find that this is an adequate approximation for the 
relatively short baselines used with ALMA so far, but a more exact 
treatment may be needed when longer baselines come into operation.

 In principle, there is a significant amount of freedom in selection of
 the weighting factors for combining the path estimates from each
 channel \citep{ALMAStirling2004}. The design work was based on
 assumption that they would be chosen to minimise the expected path
 fluctuation due the random Gaussian-like noise intrinsic to the
 radiometers and this is what the current software system implements.
 In this case the weightings factors are simply:
\begin{equation}
  \frac{1}{w_k}= \left( \delta T_{{\rm B},k}   \frac{{\rm d} L}{{\rm d} T_{{\rm
          B},k}}\right)^2 
\sum_i \left(\frac{1}{\delta T_{{\rm B},i}   \frac{{\rm d} L}{{\rm d}
      T_{{\rm  B},i}}}\right)^2
\label{eq:weights}
\end{equation}
where $\delta T_{{\rm B},k}$ is the expected intrinsic noise in the
$k$-th channel. Alternative strategies can provide slightly better
overall performance in certain conditions but at the cost of greater
effect of intrinsic noise. For example, in the case of relatively wet
weather the centre of the line is saturated and the measured sky
brightness fluctuations then correspond not only to water vapour
column and intrinsic noise fluctuations but also to the fluctuations
of the physical temperature of the water vapour layer. In this case
the measurement of path would be improved by down-weighting the
contribution of the central channel further than indicated by Equation
(\ref{eq:weights}). Such additional re-weightings are a subject
for further study and have not yet been implemented in the
production software.

The results of phase correction can often be improved slightly by
scaling down the entire correction. The reason for this is that the
intrinsic noise in the WVRs, and other sources of error in the
estimated phase correction, are usually not correlated with the true
atmospheric phase fluctuations. In more detail, the phase correction
we estimate $\hat{\delta L}$, for example according to
Equation~(\ref{eq:phase-corr-rel}), can be decomposed into the `true'
atmospheric phase change $\delta L$ and an error term $E$:
\begin{align}
  \hat{\delta L}= \delta L + E
\end{align}
If this best estimate of atmospheric phase fluctuations is applied to
correct the observed data, then the residual will be $E$ and the RMS
of the residual is simply:
\begin{align}
  \RMS{E}=\sqrt{\left<E^2\right>-\left<E\right>^2}.
\end{align}
Since we are dealing with fluctuations and error terms we can drop the
mean terms, i.e., we assume in the following that $\left<E\right>=0$
and $\left<\delta L\right>=0$. 

If, instead of correcting using the best estimate of atmospheric
fluctuations, we scale our estimate by a factor $\alpha$ before applying the correction, we find that the RMS of the residual is:
\begin{align}
  &\RMS{\delta L - \alpha \hat{\delta L}}=\RMS{(1-\alpha)\delta L -
    \alpha E}\\
  &=\sqrt{ (1-\alpha)^2\left<\delta L^2\right>  - \alpha(1-\alpha)\left<\delta L \cdot
    E\right> + \alpha^2 \left<E^2\right>}
\end{align}
In most cases, and especially when the error term
is dominated by the intrinsic thermal-like noise in the WVRs, the error term $E$ is
uncorrelated with the true atmospheric error, and therefore
$\left<\delta L \cdot E\right> \equiv 0 $.
The residual RMS then reduces to:
\begin{align}
  &\RMS{\delta L - \alpha \hat{\delta L}} = \sqrt{ (1-\alpha)^2\left<\delta
    L^2\right>  + \alpha^2 \left<E^2\right>}
\end{align}
This expression, i.e., the residual phase fluctuation, has a minimum
when $\alpha \left<E^2\right> = (1-\alpha)\left<\delta L^2\right> $,
i.e.:
\begin{align}
  \alpha_{\rm min}= \frac{\left<\delta L^2\right>}{\left<\delta
      L^2\right>+\left<E^2\right>}.
\end{align}
It can be seen from this expression that when the true atmospheric
path fluctuations dominate the noise term, $\delta L >> E$, it is best
to apply essentially the entire correction. However, when the two
terms are approximately the same $\delta L \sim E$, which is often the case on the very shortest ALMA baselines, then it is
optimal to apply \emph{only half\/} of the best estimate of the
atmospheric path fluctuations.  The optimum scaling therefore depends
on the configuration in which ALMA is observing and the atmospheric
stability. Currently this scaling is implemented as a simple
user-tunable parameter.

By default the ALMA WVRs integrate the sky observations for 1.152\,s, resulting in noise well below 0.1\,K RMS in all of the channels. This is usually sufficiently low to ensure that the resulting contribution to the phase error is small.  In very stable conditions and for short baselines it may be desirable
to smooth the observed radiometer data somewhat, i.e. use a longer integration time, in particular if the
integration period of the astronomical signal is also longer than 1.152\,s.

\subsection{Quality Control and Atmospheric Conditions Monitoring}

It is useful to derive feedback on the likely quality of the WVR
correction directly from the sky brightness temperature measurements
since this avoids the need for longer specialised observations on
quasars. We use two simple statistics to help identify any potential
problems with WVRs or with the likely efficiency of the phase
correction.

The first statistic is simply the root-mean-square fluctuation of the
estimated path from each of the WVRs. This is computed by considering
only data taken while the antennas were observing one object
(usually the science source) within a scheduling block. Additionally,
before computing the RMS, the path is corrected for the changing
airmass during the observation as the target object is tracked by the
antenna. These steps ensure that the path RMS is representative of the
actual atmospheric stability or, potentially, of any problems within
the WVR.

The second statistic that is computed is the difference between
estimates of path fluctuations obtained independently from two
channels of the WVR (the weighting and combining of channels is not
done in this case). Since the optimum phase correction coefficients of
the channels depend in a different way on the atmospheric properties,
this statistic is useful for identifying errors in the phase
correction coefficients due to wrong inference of atmospheric
properties. This statistic is also useful because any fluctuation in
sky brightness due to variable cloud cover will affect the channels in different ways, with the outer channels showing the largest effect in proportion to that of the water.

Although problems with ALMA radiometers have been rare so far,
computing these quality control statistics is important because
problems may not be apparent in astronomical data when they are
processed.

\subsection{Dealing with missing WVRs}

ALMA will need to deal with antennas which do not have functioning
WVRs.  One of the reasons for this is that ALMA consists of two arrays
of antennas: the '12m Array' which will have 50
12\,m-diameter antennas and the Atacama Compact Array (ACA) which will consist of four 12\,m-diameter and twelve 7\,m-diameter antennas. The ACA is designed to operate independently of the 12m Array
and with such short baselines it only needs basic phase
correction. It is therefore not planned for the 7\,m-diameter
antennas to have their own WVRs but rather for them to use the data from nearby
12\,m-diameter antennas.  Other possible reasons for missing WVR data are faults in the hardware or problems in transmitting the data.

The simplest approach for dealing with missing data is to interpolate
neighbouring antennas to compute a predicted WVR signal and use this
predicted WVR signal throughout the remainder of the analysis. We have
implemented interpolation as the weighted mean of signals from nearest
three antennas, with the weighting factor proportional to inverse
distance to the antenna. We find that this operates satisfactorily 
in the compact ALMA configurations since there are usually antennas with WVR measurements available that are reasonably close at hand.

\section{Initial results of phase correction}
\label{sec:phase-corr-results}

The effectiveness of phase correction can be measured by observing
known point-like sources (typically quasars) and comparing the
measured phases  to the expected values. For a point-like
source at the phase tracking centre of the interferometer, the phase
and amplitude of visibilities is expected to be constant for all
baselines and therefore any fluctuation in phases is due to a
combination of following effects:
\begin{enumerate}
  \item Atmospheric phase fluctuations.
  \item Instrumental effects.
  \item Noise in the measurement of the astronomical visibility.
  \end{enumerate}
Here we summarise the results of initial measurements of
effectiveness of radiometric phase correction, which were all made
with ALMA in compact configurations, i.e. with baseline lengths between 15 and 650\,m.

As shown in Equation~\ref{eq:phase-corr-rel} the path (and therefore
phase) correction for an antenna over short periods of time is
described by a linearised relation with the changes in each of the WVR
channels. Since the fluctuation in phase of visibility is proportional
to the \emph{difference} in path to the two antennas forming the
baseline over which the visibility is measured,
Equation~\ref{eq:phase-corr-rel} implies that there should be
proportionality between the phase fluctuations and the difference of
readings of the WVRs on the two antennas. 

\begin{figure*}
\begin{tabular}{cc}
  Channel 1 (inner-most channel) & 
  Channel 2 \\
  \includegraphics[width=0.49\linewidth]{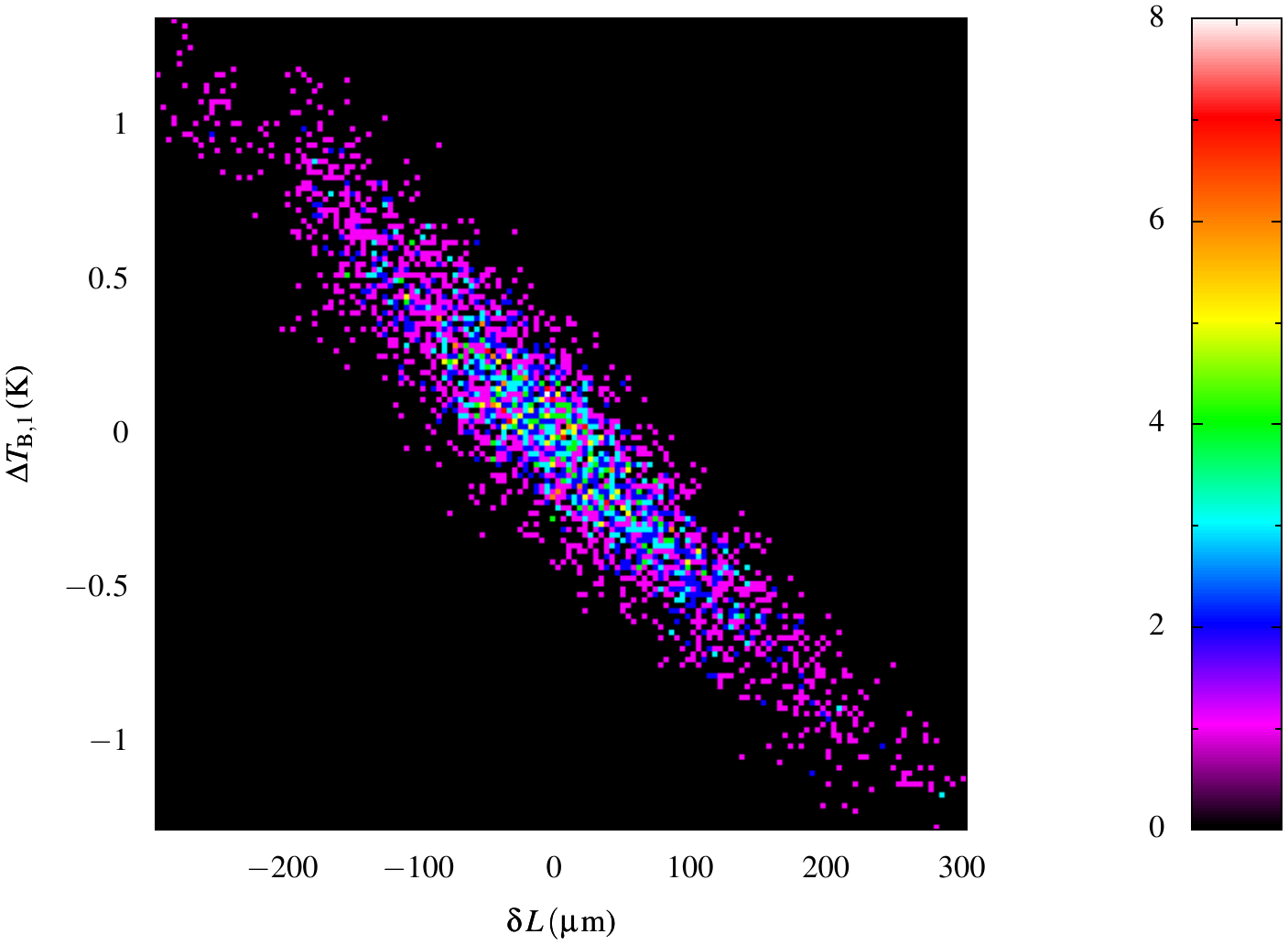}
  &
  \includegraphics[width=0.49\linewidth]{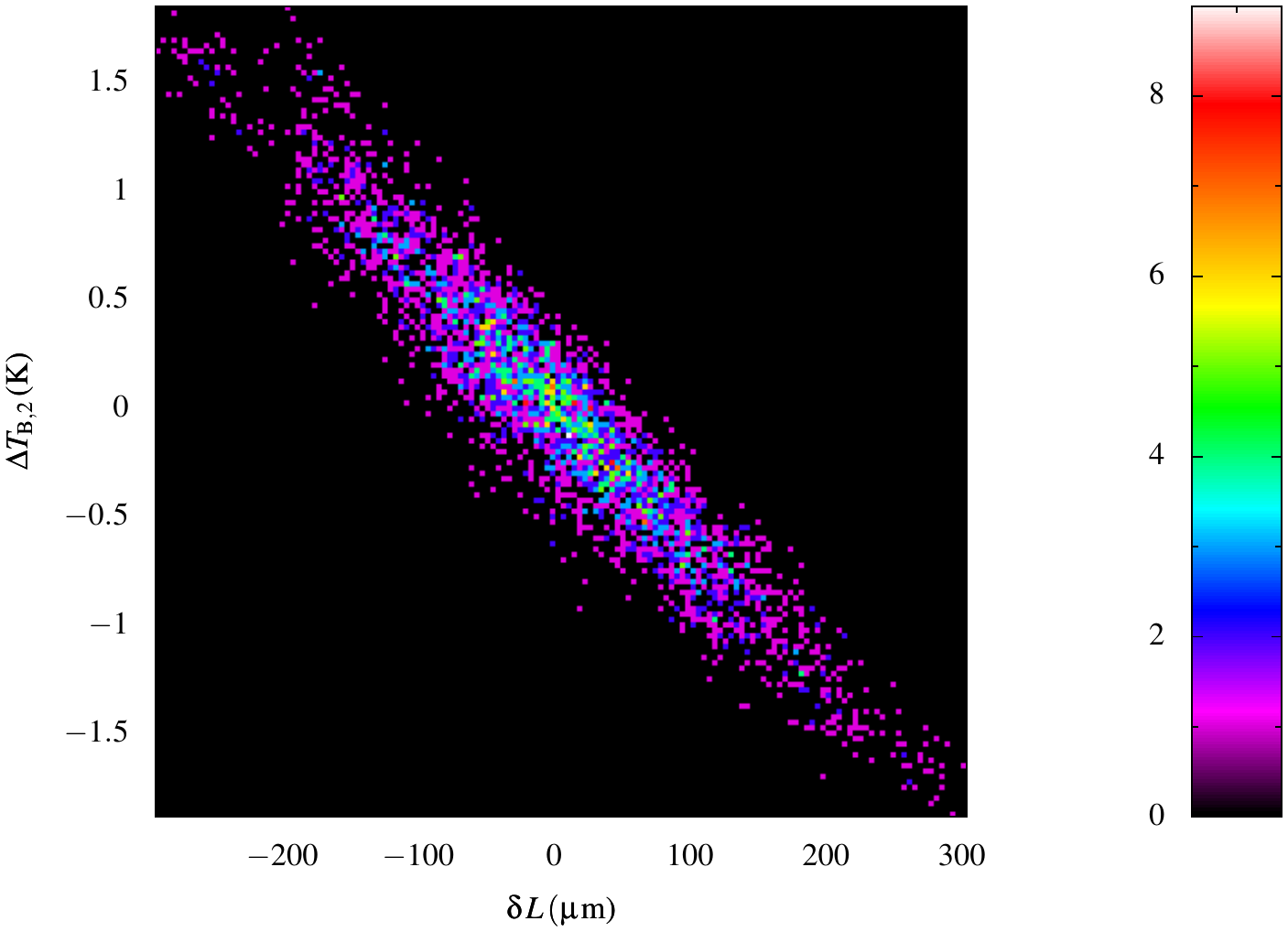}\\
  Channel 3 &
  Channel 4 (outer-most channel)\\
  \includegraphics[width=0.49\linewidth]{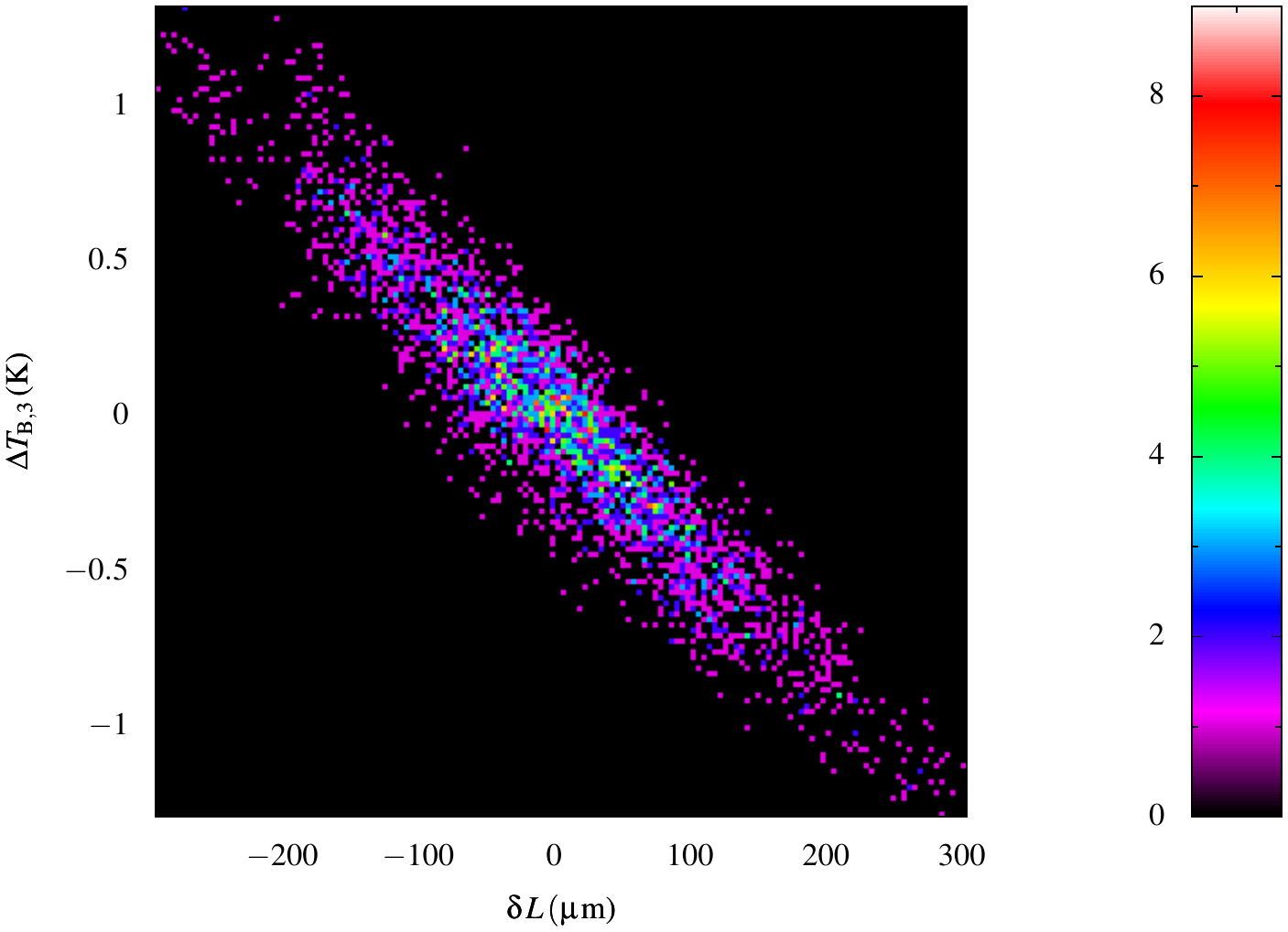}
  &
  \includegraphics[width=0.49\linewidth]{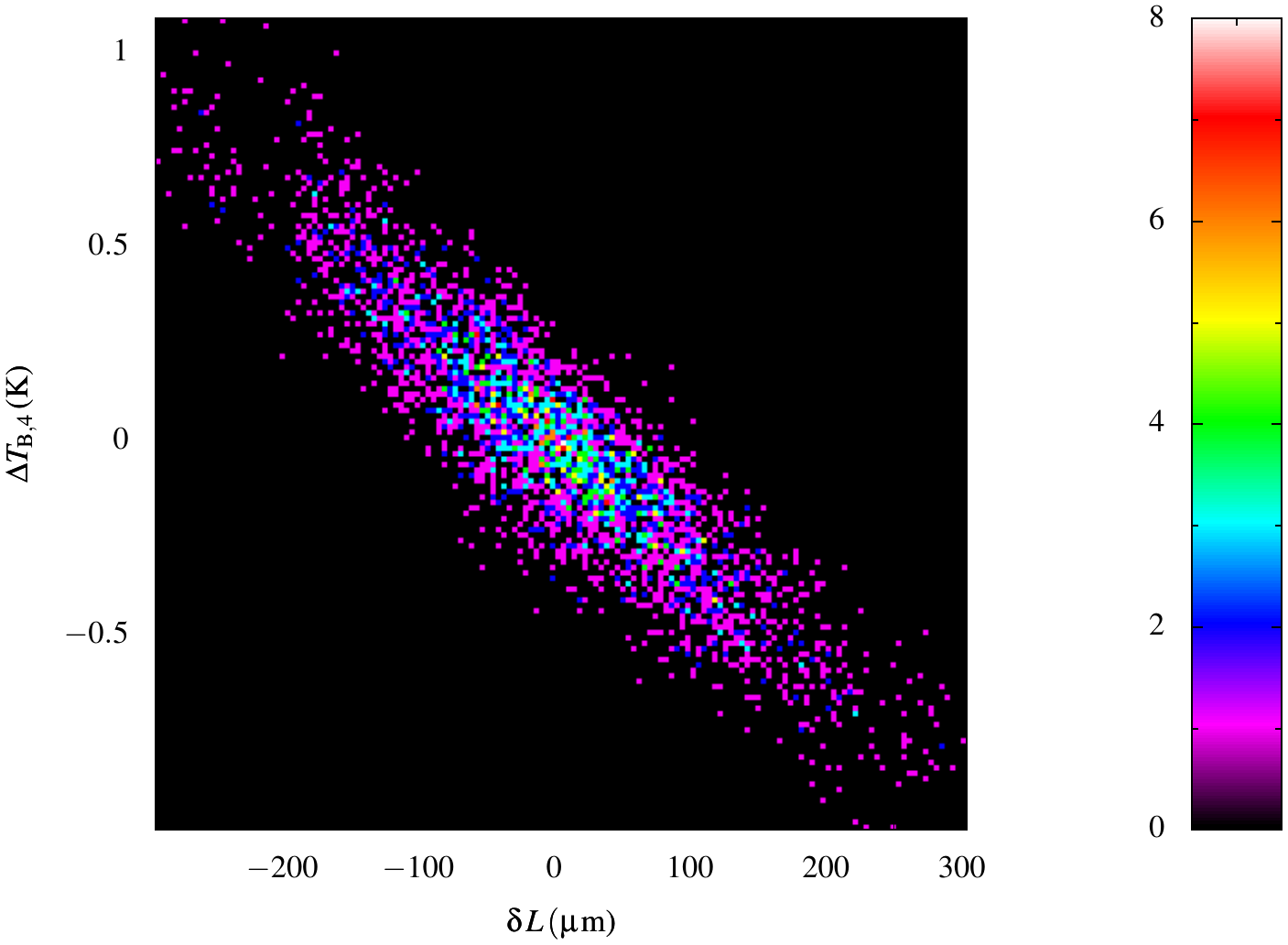}\\
\end{tabular}
\caption{Correlation between the atmospheric path error estimated
  from observations of a bright point-like object (horizontal axis)
  and the differenced WVR signal (vertical axis). Both the estimated
  path from astronomical visibilities and the WVR difference were
  filtered by removing the three minute running mean.  Each plot is a
  two-dimensional histogram where the colour scale shows how many
  points fall in each bin. The four panels correspond to the four
  channels of the radiometers.}
\label{fig:correlation}
\end{figure*}

A good way of visualising the potential efficiency of WVR phase
correction is therefore to plot the correlation between these two
quantities, i.e. the path fluctuation and the difference between WVR
measurements, as a two dimensional histogram. We show these for a
typical observation in Figure~\ref{fig:correlation}. In this figure we analyse each channel of the WVR separately and so the figure contains
four plots, one for each channel.  In each plot, the difference
between outputs of WVRs on the two antennas is used to place the data
on the vertical axis while the phase of the visibility is converted to
an apparent path fluctuation and used to place data on the horizontal
axis.  The density of points is shown by the colour.  In order to
concentrate on the changes in path over short timescales, both the
observed path fluctuation and the WVR output differences were filtered
to remove their running mean over 300\,seconds.

It can be seen in Figure~\ref{fig:correlation} that the correlation
between the WVR differences and the measured path fluctuations is very
tight. A formal regression analysis shows that the adjusted $R^2$
values are 0.82--0.9, depending on the channels, and the RMS residuals in terms of brightness temperature differences are 0.13--0.19\,K. This can be compared to
expected RMS residuals of 0.1--0.14\,K due to thermal noise in the
WVRs only.  Furthermore it can be seen that there are no out-lying points which would be seen if there were spikes or
glitches in either the WVR measurements or the interferometric phases.

\begin{figure}
    \begin{tabular}{c}
      \includegraphics[width=0.95\columnwidth]{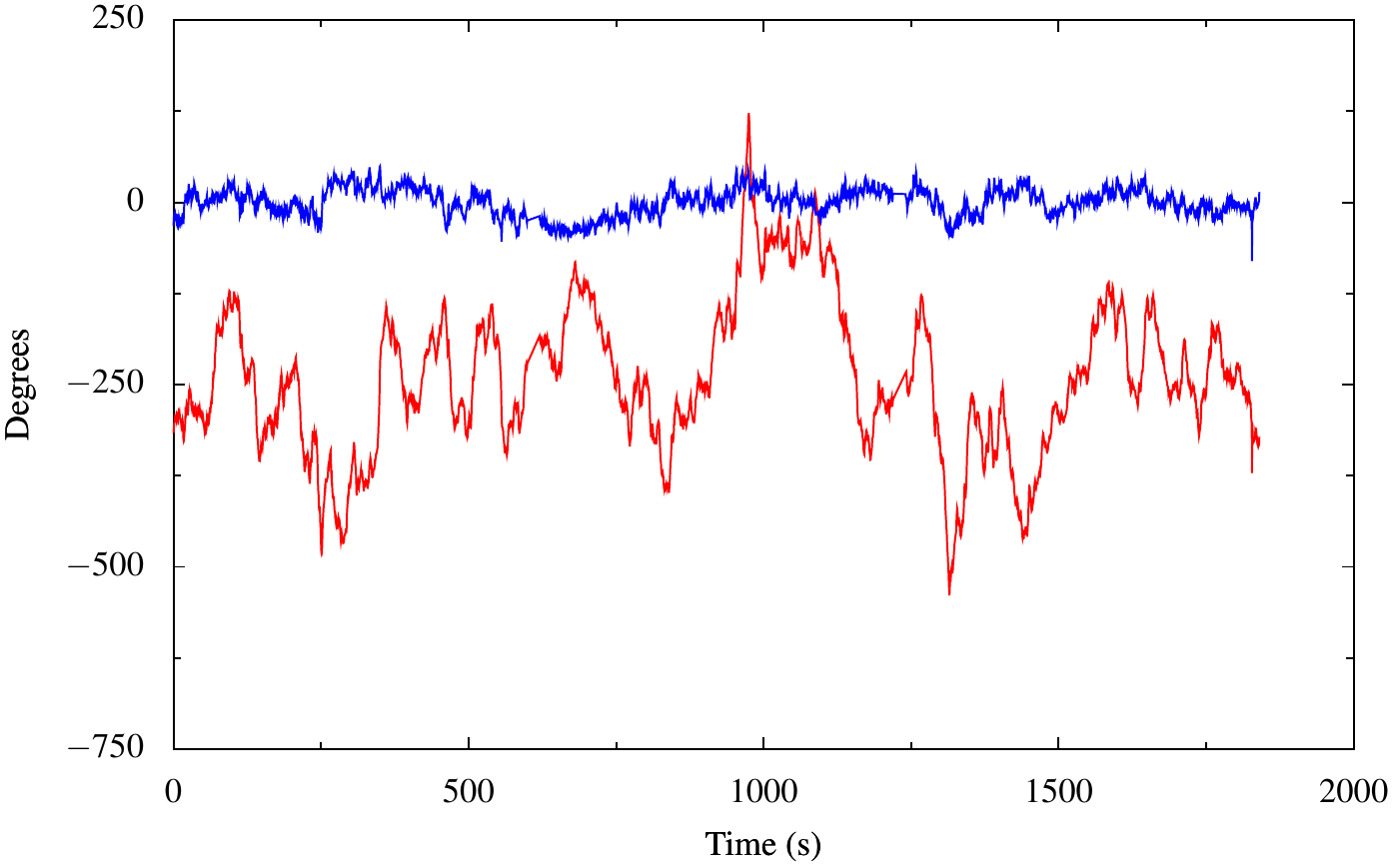}
    \end{tabular}
    \caption{Test observation at 90\,GHz of a strong quasar on a
      $\sim650\,$m baseline with ALMA. The red line is the phase of
      the observed (complex) visibility on this baseline -- note that
      for a quasar (or other point-like) source at the tracking centre
      of the interferometer we expect a constant phase in time. The
      blue line is the visibility phase after correction of the data
      based on the WVR signals and using the {\tt wvrgcal} program.}
    \label{fig:result-daytime}
  \end{figure}

\begin{figure}
    \begin{tabular}{c}
      \includegraphics[width=0.95\columnwidth]{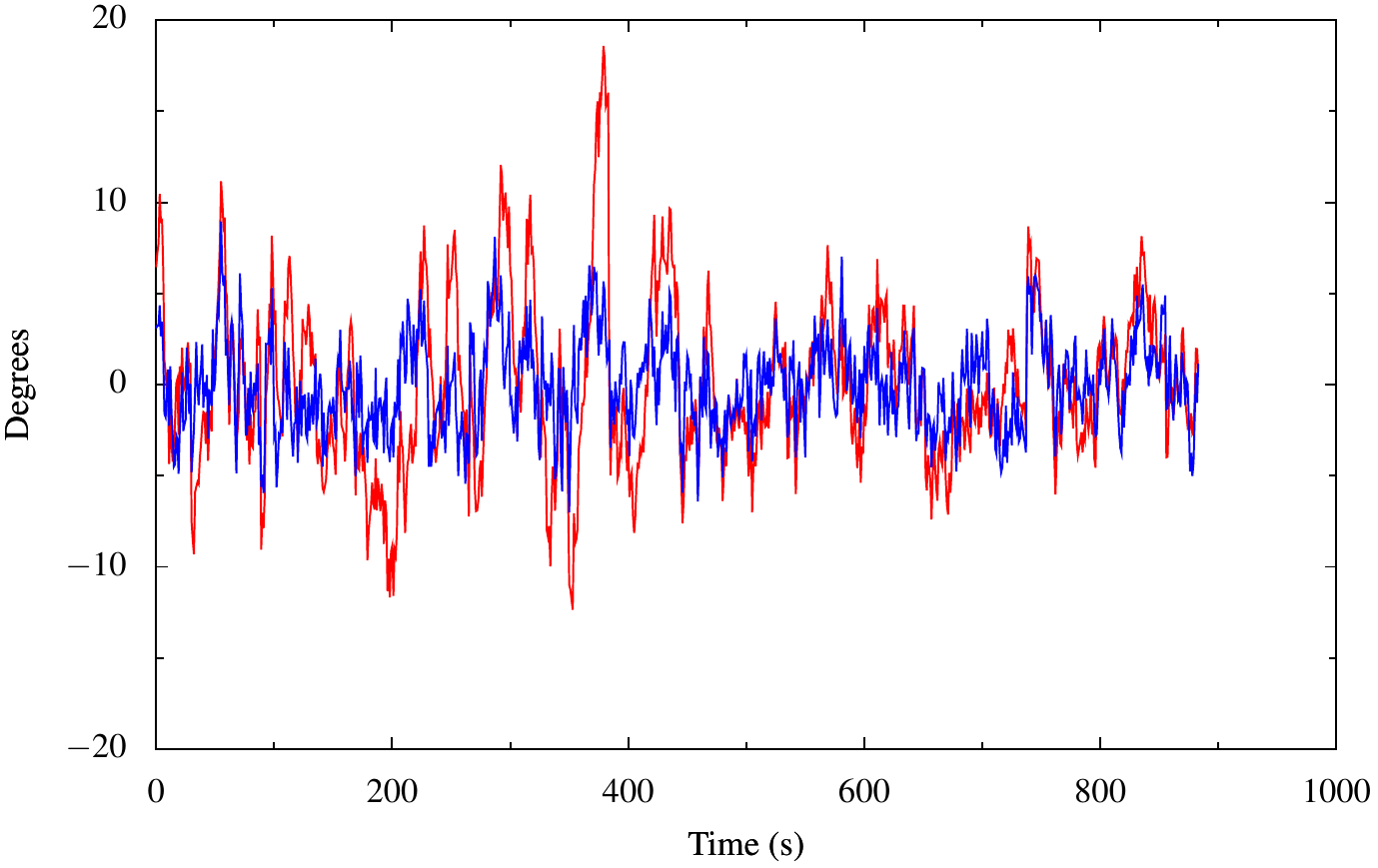}
    \end{tabular}
    \caption{Test observation on a short ($\sim 25\,$m) baseline in
      excellent weather conditions.  The phase of the uncorrected
      astronomical visibility is again in red while the phase after WVR
      correction is in blue. Note the much smaller range of the vertical axis compared to Figure ~\ref{fig:result-daytime}.}
    \label{fig:result-best}
\end{figure}

The improvement in phase stability of astronomical visibility after
WVR correction is illustrated in Figures~\ref{fig:result-daytime} and
\ref{fig:result-best}. The first figure
(Figure~\ref{fig:result-daytime}) shows the improvement in relatively
humid conditions (PWV $\sim 2.2\,$mm) during unstable daytime
conditions on baseline which is about 650\,m long. As can be seen the
phase fluctuations are reduced by almost an order of magnitude (in
terms of path RMS from about 1\,mm to 0.16\,mm), meaning that they are
reduced from a level which would almost completely decorrelate the
signal to one which is adequate for interferometric imaging.

The results in the dry (PWV $\sim 0.5$\,mm) and stable night-time
conditions on a short baseline of $\sim 20\,$m are shown in
Figure~\ref{fig:result-best}. In this example the path fluctuation
(after removing the 3\,minute running mean) is reduced from
14\,\micron\ to 7\,\micron. Although the relative improvement is much
smaller than that shown in Figure~\ref{fig:result-daytime}, this
example illustrates the very high absolute level of performance that
can be achieved.  Both examples also illustrate the good long-term
stability of the WVRs with no noticeable drift.

The initial testing has confirmed the presence of some shortcomings
which were, to some extent, anticipated during the design stage. The
most obvious of these is the effect of cloud on the phase correction.
Variable cloud cover causes fluctuations in sky brightness which do
not have the same relation to the path fluctuation as those due to
water vapour and this leads to erroneous phase correction. Even on
relatively short baselines this effect has been found the be large
enough to make the phases after correction \emph{worse\/} than before
correction on some occasions. While sky brightness measurements from the outermost channel 
have been useful in identifying when clouds are making a significant 
contribution to sky brightness, we have not yet found a way of using 
these measurements in a way that improves the phase correction in the 
presence of clouds.

The second shortcoming which has been identified is the presence of  residual phase
errors which are not correlated with the water vapour signal. The
residual errors only become noticeable in quite dry conditions and
their magnitude appears to increase with increasing length of the baseline. The origin of these residual errors has not yet been
identified conclusively but it seems likely they are due to the
fluctuations of refractive index of the dry component of the air, i.e. density fluctuations.  Such density fluctuations will arise from temperature variations, especially in the daytime, but there may also be a component due to the dynamical pressure changes caused by the wind. 

An important aspect of phase correction which we have not 
examined in this paper is the performance of the correction in
combination with phase referencing. In this case the WVR corrections 
can first be applied to the whole sequence of data, consisting of 
interleaved observations of the astronomical source and the calibrator, 
and a further correction, based on the remaining phase variations seen 
in the calibrator, can then be applied to the astronomical data.  This 
combination of the two corrections should result in better imaging 
quality than the application of either of the corrections on its own, 
since it addresses both the short-term atmospheric fluctuations and 
the somewhat longer-term phase drifts in the interferometer system.
Phase referencing can however involve a significant change in telescope 
elevation. This might  lead to instrumental effects in the 
WVRs which would cause errors in their readings.  More significantly 
the changes in the observed sky brightness (due to change in airmass)
may be quite large. Although the changes in brightness will be  
common for all the antennas in the array, small differences in the 
factors  on the different WVRs that convert from sky brightness to 
excess path will produce errors in the phase corrections. Such 
elevation-dependent errors can mimic other sources of error, such as 
those due to inaccurately determined positions of the telescopes. We 
have not yet been able to make a proper assessment of the errors, 
where we can separate the various contributions, in the case of 
phase-referenced observations. In practice, the impact of such errors 
on ALMA observations can often be reduced by using self-calibration 
on timescales which are long enough to provide adequate signal to noise 
ratio but short enough to correct for the slowly changing 
elevation-dependent errors. Additionally, the sensitivity of ALMA is
increasing as more telescopes are commissioned into the array and
this means fainter (and therefore closer) phase calibrators can be
used, which reduces any elevation-dependent errors. The accuracy of 
phase correction when combined with phase referencing is nevertheless 
an important topic which we hope to examine as part of future work.

\section{Summary}

We have described the design of the ALMA water-vapour radiometer phase-correction system. Compared to most previous systems, the ALMA system has the following advantages:
\begin{enumerate}
\item It observes the very strong 183\,GHz water vapour line which has a
  phase correction coefficient (i.e., the relationship between sky
  brightness change and electrical path change) as high as 40\,K/mm
  \item Internal continuous two-load calibration
  \item Optical and mechanical design optimised during telescope
    design stages, giving low spill-over and rigid telescope mounting
\end{enumerate} 
Additionally ALMA has the advantage that it is at a higher and dryer site,
leading to intrinsically smaller water vapour fluctuations, and that
its sensitivity is large, which means that phase calibration sources
can often be found close to science targets.

The successes of this system to date include:
\begin{enumerate}
\item Initial tests of the phase correction system have shown that the radiometers 
  have excellent sensitivity and stability;
\item Significant and often dramatic improvement in phase
    stability have been achieved under most conditions;
\item Phase correction is already in routine use for ALMA
    science observing.
  \end{enumerate}
  
The main limitations are erroneous phase corrections when there is thick
cloud cover and the presence of some residual phase errors which may
be due to air-density fluctuations.

\section*{Acknowledgements}

This work was supported in part by the European Commission's Sixth
Framework Programme as part of the wider `Enhancement of Early ALMA
Science' project. We would like to thank the referee Dr Robert
  Lucas for helpful and prompt suggestions for improving the paper.

\bibliographystyle{aa} 
\bibliography{alma.bib}

\end{document}